\documentclass[11pt]{article}
\usepackage{jheppub,amsmath, amsthm, amssymb,slashed,url}
\usepackage{graphicx}
\usepackage{epstopdf}

\usepackage{csquotes}
\usepackage{footnote}
\makesavenoteenv{tabular}
\makesavenoteenv{table}

\def\OMIT#1{{}}

\newcommand{\bit}[1]{\mbox{\boldmath$#1$}}

\def\OMIT#1{{}}

\def\ga{g_{\scriptscriptstyle A}}

\def\yo2{{f_\pi^2}}
\def\llra{{\relbar\joinrel\rightarrow}}
\def\mapright#1{{\smash{\mathop{\llra}\limits_{#1}}}}

\def\oneht{\textstyle{1\over 2} }

\newcommand{\BE}{\begin{equation}}
\newcommand{\EE}{\end{equation}}
\newcommand{\BA}{\begin{eqnarray}}
\newcommand{\EA}{\end{eqnarray}}

\font\teneurm=eurm10 \font\seveneurm=eurm7 \font\fiveeurm=eurm5
\newfam\eurmfam
\textfont\eurmfam=\teneurm \scriptfont\eurmfam=\seveneurm
\scriptscriptfont\eurmfam=\fiveeurm

\font\teneusm=eusm10 \font\seveneusm=eusm7 \font\fiveeusm=eusm5
\newfam\eusmfam
\textfont\eusmfam=\teneusm \scriptfont\eusmfam=\seveneusm
\scriptscriptfont\eusmfam=\fiveeusm

\font\tencmmib=cmmib10 \skewchar\tencmmib='177
\font\sevencmmib=cmmib7 \skewchar\sevencmmib='177
\font\fivecmmib=cmmib5 \skewchar\fivecmmib='177
\newfam\cmmibfam
\textfont\cmmibfam=\tencmmib \scriptfont\cmmibfam=\sevencmmib
\scriptscriptfont\cmmibfam=\fivecmmib

\def\Pi{\varPi}

\usepackage{marginnote}
\usepackage{subcaption}
\newcommand{\gpin}{\ensuremath{g_{\pi N}}}
\newcommand{\fpi}{\ensuremath{F_\pi}}
\newcommand{\mn}{\ensuremath{m_N}}
\newcommand{\mpi}{\ensuremath{M_\pi}}

\newcommand{\ompicubed}{\ensuremath{\mathcal{O}(\mpi^3)}}

\newcommand{\awd}{\ensuremath{\Delta_{AW}}}

\dedicated{NT@UW-16-04}

\title{Chiral corrections to the Adler-Weisberger sum rule}

 \author{Silas R.~Beane and Natalie Klco}

\affiliation{Department of Physics, University of Washington,
Seattle, WA 98195}

\date{\mydate}

\abstract{The Adler-Weisberger sum rule for the nucleon axial-vector
  charge, $\ga$, offers a unique signature of chiral symmetry and its
  breaking in QCD. Its derivation relies on both algebraic aspects of
  chiral symmetry, which guarantee the convergence of the sum rule,
  and dynamical aspects of chiral symmetry breaking---as exploited
  using chiral perturbation theory---which allow the rigorous
  inclusion of explicit chiral symmetry breaking effects due to
  light-quark masses. The original derivations obtained the sum rule
  in the chiral limit and, without the benefit of chiral perturbation
  theory, made various attempts at extrapolating to non-vanishing pion
  masses. In this paper, the leading, universal, chiral corrections to
  the chiral-limit sum rule are obtained. Using PDG data, a recent
  parametrization of the pion-nucleon total cross-sections in the
  resonance region given by the SAID group, as well as recent
  Roy-Steiner equation determinations of subthreshold amplitudes,
  threshold parameters, and correlated low-energy constants, the
  Adler-Weisberger sum rule is confronted with experimental data.
  With uncertainty estimates associated with the cross-section
  parameterization, the Goldberger-Treimann discrepancy, and the
  truncation of the sum rule at $\mathcal{O}(\mpi^4)$ in the chiral
  expansion, this work finds $\ga = 1.248 \pm 0.010 \pm 0.007 \pm
  0.013$.}

\begin{document} \maketitle

\section{Introduction}
\label{intro}

\noindent The success of the Adler-Weisberger (AW) sum
rule~\cite{Adler:1965ka,Weisberger:1965hp} in calculating the nucleon
axial-vector charge, $g_A$, was important
historically~\cite{Adler:2006if} as it provided a striking pre-QCD
confirmation of the importance of chiral symmetry in understanding
nucleon structure through the strong interaction. The original derivation of the sum rule used some
of the language of the infinite momentum frame as well as
then-available knowledge of current algebra low-energy
theorems\footnote{For a detailed description of these methods, see
  Ref.~\cite{deAlfaro:1973zz}}. These two technologies have substantially
advanced and evolved, and therefore it is interesting to reassess the
theoretical basis for the AW sum rule. In addition, knowledge of the
experimental total cross-sections in the resonance region~\cite{Workman:2012hx}---which is
essential for a confrontation of the sum rule with experiment---as
well as overall knowledge of the pion-nucleon interaction~\cite{Hoferichter:2015tha} have
advanced to a high level. Therefore, an updated analysis of the
experimental validity of the AW sum rule and its implications for the
nucleon axial-vector charge, with controlled uncertainties, is timely.

It is worth summarizing the standard view of how the AW sum rule is
obtained.  Firstly, soft-pion theorems are derived using current
algebra methods or chiral perturbation
theory~\cite{Weinberg:1978kz,Gasser:1983yg,Gasser:1987rb,Jenkins:1990jv,Becher:2001hv}
($\chi$PT) to obtain the crossing-odd, forward scattering amplitude at
a special low-energy kinematical point. The Regge model of asymptotic
behavior is then invoked to argue that this amplitude vanishes
sufficiently quickly at high energy to guarantee an unsubracted
dispersion relation, and the optical theorem is used to replace the
absorptive part of the scattering amplitude with the total
cross-section. While there is nothing wrong with this perspective of
the sum rule, one goal of this paper is to stress that it is not
necessary to invoke Regge lore in deriving the AW sum rule~\cite{deRafael:1997ea}, as the
scattering amplitude in question is explicitly calculable in the Regge
limit ($s\gg -t$), and is found to vanish as a consequence of the
chiral symmetry of QCD~\cite{Beane:2015ufo,Weinberg:1969hw}. The
convergence of the AW sum rule is therefore a direct consequence of
the chiral symmetry of QCD and does not depend on model input.

In the original derivations, the major theoretical hurdle in
confronting the AW sum rule with experiment was the ambiguity in
extrapolating from the world of massless pions to the physical
world~\cite{Weinberg:1971}, as $\chi$PT did not yet exist. Here, the
leading chiral corrections to the chiral-limit expression of the AW
sum rule are obtained. Of course, these chiral corrections are
universal. However, there is no unique analog of the AW sum rule away
from the chiral limit, as there is freedom to evaluate the underlying
dispersion relation at the threshold point, or in the subthreshold
region, in such a way that the resulting sum rule reduces to the AW
sum rule in the chiral limit. In the language of effective field
theory, these variants are equivalent, up to distinct resummations of
pion-mass effects.  It is natural to formulate the AW sum rule in a manner
that leaves the chiral-limit form invariant and includes chiral
corrections perturbatively using $\chi$PT. This sum rule can then be
treated as a constraint on $g_A$ that is rigorous in QCD up to
subleading corrections in the chiral expansion.

This paper is organized as follows, Section~\ref{NandC} introduces the
basic pion-nucleon scattering conventions that are essential for our
investigation. Section~\ref{ABCS} reviews the connection between
algebraic chiral symmetry and the soft asymptotic behavior of the
crossing-odd, forward pion-nucleon scattering amplitude.  In
Section~\ref{MDR}, the well-known, crossing-odd, forward dispersion
relation is written down and evaluated at several kinematical
points. While the results of this section are well
known, they are essential for what follows. The leading chiral
corrections to the chiral-limit form of the AW sum rule are derived in
Section~\ref{AWD}. A confrontation of the AW sum rule with experimental
data requires detailed knowledge of the total pion-nucleon
cross-sections.  Therefore, a parametrization of the cross-sections
across all relevant ranges of energies is constructed in Section~\ref{CALC}
and used to put the AW sum rule to the test.  Finally, we state our
conclusions in Section~\ref{conc}.

\section{Notation and conventions}
\label{NandC}

\noindent We use the standard conventions of Ref.~\cite{Hohler:1983}.
The four momenta of the incoming nucleon and pion are $p$ and $q$ and
the four momenta of the outgoing nucleon and pion are $p'$ and $q'$.
Therefore, $s=(p+q)^2$, $t=(q-q')^2$ and $u=(p-q')^2$ with
$s+t+u=2M_\pi^2+2m_N^2$.  The lab energy of the incoming pion is
$\omega={(s-m_N^2-M_\pi^2)}/{2m_N}$ and the lab momentum of the
incoming pion is $k=\sqrt{\omega^2-M_\pi^2}$.  It is convenient to
express the energy in terms of the crossing-symmetric variable
$\nu={(s-u)}/{4 m_N}$. In the forward limit, $\nu =\omega$.  We denote
the chiral limit values of $g_A$, $F_\pi$, $m_N$ and $M_\pi$ as $g$,
$F$, $m$ and $M$. The scattering amplitude can be expressed as
\BA
T_{\alpha\beta} & =&  \delta_{\alpha\beta}T^+ \ +\ \oneht [{\bf \tau}_{\alpha},{\bf \tau}_\beta]\,T^- \ ; \\
T^{\pm}& =&\bar{u}(p')\left\{ D^{\pm}(\nu, t)-\frac{1}{4 m_N} [\,q\hspace{-0.5em}/^{\,\prime},\, q\hspace{-0.5em}/\hspace{0.15em}]B^{\pm}(\nu,t) \right\}u(p) \ ,
\label{eq:sadefs}
\EA
where $\alpha,\beta$ are isospin indices. This paper is about crossing-odd, forward-scattering and therefore concerns itself solely with $D^-(\nu, 0)$,
which is related to the total pion-proton ($\pi p$) scattering cross-sections via the optical theorem:
\BA
\mbox{Im} D^-(\nu,0)\ =\ k\,\sigma^-(\nu) \ =\ k \oneht \left( \sigma^{\pi^-p}(\nu) \ -\ \sigma^{\pi^+p}(\nu)\right) \ .
\label{eq:opticalT}
\EA
As crossing symmetry implies that $D^-(\nu,0)/\nu$ is even in $\nu$, the expansion of the amplitude
about $\nu=0$ in the forward direction is
\BA
\frac{D^-(\nu,0)}{\nu}\ =\ \frac{g_{\pi N}^2}{m_N}\,\frac{\nu_B}{\nu_B^2-\nu^2}\ - \ \frac{g_{\pi N}^2}{2 m_N^2} \ +\  d^-_{00} \ +\ d^-_{10}\,\nu^2 \ +\ \ldots \ ,
\label{eq:STexp}
\EA
where $\nu_B \equiv -{M_\pi^2}/{2\,m_N}$, $g_{\pi N}$ is the pion-nucleon coupling constant, and the $d^-_{n0}$ are subthreshold amplitudes.
The scattering length $a_{0+}^-$ is defined via
\BA
4\pi a_{0+}^-\,\left( 1\,+\,\frac{M_\pi}{m_N} \right) \ \equiv \ D^-(\nu,0)|_{\nu =M_\pi}\ .
\label{eq:sldefined}
\EA

It will prove useful to give the chiral expansions of various quantities~\cite{Becher:2001hv,Hoferichter:2015hva}.
The pion-nucleon coupling constant may be expressed as
\BA
g_{\pi N}\ =\ \frac{g_A m_N}{F_\pi}\left(\; 1\; +\; \Delta_{GT}\; \right) \ ,
\label{eq:gpin}
\EA
where $\Delta_{GT}$ is the Goldberger-Treiman (GT) discrepancy~\cite{Goldberger:1958tr,Bernard:1992qa}, whose chiral
expansion is
\BA
\Delta_{GT}\ =\ -\frac{2 \bar{d}_{18}M^2}{g}\ +\  \mathcal{O}(M^4) \ .
\label{eq:gtd}
\EA
The chiral expansion of the leading subthreshold amplitude is~\cite{Fettes:2000xg}
\BA
d_{00}^- &=&\frac{1}{2{F_\pi}^2}+\frac{4\,( \bar{d}_1+\bar{d}_2 +
      2\,\bar{d}_{5} ) \,
      {M_\pi^2}}
{F_\pi^2}  +
   \frac{{g_A^4}\,M_\pi^2}
    {48\,\pi^2\,F_\pi^4} \nonumber \\
&& \qquad\qquad -\ M_\pi^3\left(\frac{8+12g_A^2+11g_A^4}{128\pi F_\pi^4 m_N}-\frac{4c_1+g_A^2(c_3-c_4)}{4\pi F_\pi^4}\right) \ +\  \mathcal{O}(M_\pi^4) \ ,
\label{eq:d00exp}
\EA
where the ${c}_i$ and $\bar{d}_i$ in Eqs.~\eqref{eq:gtd} and \eqref{eq:d00exp} are (scale-independent) low-energy constants (LECs) that
are unconstrained by chiral symmetry.

\section{Asymptotic behavior and chiral symmetry}
\label{ABCS}

\noindent The existence of a sum rule hinges on the asymptotic behavior of the
crossing-odd forward amplitude. As mentioned above, this amplitude is
special in QCD as its asymptotic behavior is constrained by chiral
symmetry. This constraint is most easily derived by considering the
light-cone current algebra that naturally arises when QCD is quantized
on light-like hyperplanes. Remarkably, there is a set of scattering
amplitudes whose Regge-limit values can be expressed as matrix
elements of the current
algebra moments~\cite{Beane:2015ufo,Weinberg:1969hw}. The Regge-limit value of
the crossing-odd, forward, $\pi p$ scattering amplitude is given
by~\cite{Beane:2015ufo}
\BA
\frac{D^-(\nu,0)}{\nu}|_{\nu =\infty}  = \frac{1}{F_\pi^2}
\int \frac{d k^+ d^2 \bit{k}_\perp}{2 k^+ (2 \pi)^3}
\langle\, {\rm p}\,,\,\lambda\,;\,k\, |\left( 2{\tilde {Q}}^3 - {[\, {\tilde {Q}}^+_{5}(x^+) \, ,\, {\tilde {Q}}^-_{5}(x^+)\, ]} \right)|\, {\rm p} \,,\,\lambda\,;\,k\,\rangle \ ,
\label{eq:DminusCS}
\EA
where $p$ denotes the proton, $k=(k^+,\bit{k}_\perp)$ is the
null-plane momentum, and ${\tilde Q}^\pm_5(x^+)\equiv {\tilde
  Q}^1_5(x^+)\pm i {\tilde Q}^2_5(x^+)$ with ${\tilde
  Q}^\alpha_5(x^+)$ the null-plane axial-vector
charge~\cite{Beane:2015ufo}.  The conserved null-plane vector charge
is ${\tilde Q}^\alpha$. The null-plane axial-vector charges are not
conserved, even in the chiral limit, and therefore they carry explicit
dependence on null-plane time, $x^+$.  This property allows the
charges to mediate transitions between states of different energies,
and is, in a fundamental sense, responsible for the existence of the
AW sum rule, as will be further discussed below. As QCD with two
massless flavors has an $SU(2)_L\otimes SU(2)_R$ invariance, for any
initial quantization surface, there exist charges satisfying the
associated Lie algebra. In particular, if one works with null planes
then the following Lie bracket is clearly satisfied at the operator
level:
\BA
[\, {\tilde Q}_{5\alpha}(x^+)\, ,\, {\tilde Q}_{5\beta}(x^+)\, ]\, =\,  i\,\epsilon_{\alpha\beta\gamma}\,  {\tilde Q}_\gamma \ ,
\label{eq:npchargealgA}
\EA
which guarantees, via Eq.~(\ref{eq:DminusCS}), the vanishing
asymptotic behavior of the crossing-odd, forward $\pi p$ scattering
amplitude~\footnote{ This soft asymptotic behavior is consistent with
  the Regge model which suggests ${D^-(\nu,0)}/{\nu}\
  \mapright{\nu\rightarrow\infty} \ \nu^{\alpha_\rho(0)-1}$ with
  $\alpha_\rho(0)\sim 0.5$. }.  In the chiral limit, the AW sum rule
then follows either through direct evaluation of the matrix element of
the Lie bracket of
Eq.~(\ref{eq:npchargealgA})~\cite{deAlfaro:1973zz,Beane:2015ufo} or by
using dispersion theory (see below), and is given by
\BA
g^2 \ =\ 1 \ -\ {{2F^2}\over\pi}\int_{0}^\infty {{d\nu}\over\nu}\;
\Big\lbrack\sigma^{\pi^- p}(\nu )-\sigma^{\pi^+ p}(\nu )\Big\rbrack \ ,
\label{eq:AWcl}
\EA
where it is understood that the cross-section in the integrand is evaluated from the chiral-limit amplitude.
Replacing all chiral-limit parameters and amplitudes with the physical ones yields a sum rule that
can be confronted with experiment:
\BA
g_A^2 \ =\ 1 \ -\ {{2F_\pi^2}\over\pi}\int_{M_\pi}^\infty {{d\nu}\over\nu^2}\;k
\Big\lbrack\sigma^{\pi^- p}(\nu )-\sigma^{\pi^+ p}(\nu )\Big\rbrack \ .
\label{eq:AWnaive}
\EA
Of course this sum rule is valid only to $\mathcal{O}(M_\pi^0)$ and receives a
non-trivial correction at each order in the chiral expansion. It is the main purpose of this
paper to compute the leading chiral corrections and confront the corrected sum rule
with data.

\section{Sum rule review}
\label{MDR}

\subsection{Crossing-odd forward dispersion relation}
\label{general}

Away from the chiral limit, the asymptotic behavior of the crossing-odd, forward scattering amplitude guaranteed
by the chiral symmetry algebra is unchanged~\footnote{This claim rests on the simple observation that turning on light-quark masses with $m_u,m_d\ll \Lambda_{QCD}$ does not alter the
asymptotic behavior of scattering amplitudes when $s\gg \Lambda^2_{QCD}$.  Note that throughout this paper only $\chi$PT with two light flavors is pertinent.} and therefore the scattering amplitude satisfies the dispersive representation
\BA
\frac{D^-(\nu,0)}{\nu} \ = \ \frac{g_{\pi N}^2}{m_N} \frac{\nu_B}{\left(\nu_B^2\,-\,\nu^2 \right)}\ +\ \frac{2}{\pi}P \int \frac{{\rm Im} D^-(\nu',0)d\nu'}{\nu^{\prime 2}-\nu^2} \ ,
\label{eq:MasterDispersion}
\EA
where $P$ denotes the principal value. Apart from general physical
principles, the sole ingredient that enters the derivation of
Eq.~(\ref{eq:MasterDispersion}) is the asymptotic behavior implied by
chiral symmetry via Eq.~(\ref{eq:DminusCS}) and
Eq.~(\ref{eq:npchargealgA}). In the chiral limit, this dispersion
relation is profitably exploited only at threshold, $\nu_{th}=0$,
which leads to Eq.~(\ref{eq:AWcl}) using the formulas of
Section~\ref{NandC}. However, away from the chiral limit, both the
threshold point, $\nu_{th}=M_\pi$, and the subthreshold point,
$\nu=0$, provide useful sum rules.

\subsection{Threshold evaluation}
\label{te}

\noindent Evaluating the general dispersion relation, Eq.~\eqref{eq:MasterDispersion}, at $\nu_{th}=M_\pi$ gives the sum rule
\BA
a_{0+}^-\,\left( 1\,+\,\frac{M_\pi}{m_N} \right) \ = \ \frac{g_{\pi N}^2}{2\pi} \frac{M_\pi}{\left(4\mn^2\,-\,M^2_\pi \right)}\ +\
\frac{M_\pi}{4\pi^2}\int_{M_\pi}^\infty \frac{k \Big\lbrack\sigma^{\pi^- p}(\nu)-\sigma^{\pi^+ p}(\nu)\Big\rbrack d\nu}{\nu^{ 2}-M_\pi^2} \ .
\label{eq:GMO}
\EA
Eq.~(\ref{eq:GMO}) is the Goldberger-Miyazawa-Oehme (GMO) sum
rule~\cite{Goldberger:1955zza} which predates the AW sum rule. Note
that the GMO sum rule follows only from the asymptotic constraint of
Eq.~(\ref{eq:DminusCS}).  Therefore, while this sum rule is a
consequence of the chiral symmetry algebra, it has nothing to do with
$\chi$PT unless one chooses to expand the various physical quantities
that enter the sum rule in the chiral expansion.  Recent analyses of
this sum rule can be found in Refs.~\cite{Abaev:2007nq,Baru:2010xn,Baru:2011bw}.

\subsection{Subthreshold evaluation}
\label{ste}

\noindent Evaluating the general dispersion relation at the subthreshold point, $\nu=0$, gives the sum rule~\cite{Becher:2001hv,Hohler:1983}
\BA
d^-_{00} =  \frac{\gpin^2}{2\mn^2}\ +\ \frac{1}{\pi}  \int_{\mpi}^\infty \frac{d\nu}{\nu^{ 2}} {k \left[ \sigma^{\pi^-p}(\nu) \ -\ \sigma^{\pi^+p}(\nu)\right] } \ .
\label{eq:d00sumrule}
\EA
Again, this sum rule relies solely on chiral symmetry to validate the soft asymptotic behavior of the cross-section.

\subsection{Higher moments}
\label{Hm}

\noindent There are also sum rules that follow from the higher moments ($n>0$) of the general dispersion relation, Eq.~(\ref{eq:MasterDispersion}), around $\nu=0$:
\BA
d_{n0}^- = \frac{1}{\pi} \int_{\mpi}^\infty \frac{d\nu}{\nu^{ 2(n+1)}} {k\left[ \sigma^{\pi^-p}(\nu) \ -\ \sigma^{\pi^+p}(\nu)\right]} \ .
\label{eq:momentsumrulesODD}
\EA
These moment sum rules are not related to chiral symmetry as they rely
solely on unitarity via the Froissart-Martin
bound~\cite{Froissart:1961ux,Martin:1962rt}, which requires
$\sigma(\nu)< \ln^2\nu$ at large $\nu$ (See also
Ref.~\cite{Hohler:1983}). These moments will prove to be useful checks
of the parametrization of the total cross-section that is developed
below.

\section{The AW discrepancy}
\label{AWD}

\noindent The chiral corrections to the (chiral limit) AW sum rule of
Eq.~\eqref{eq:AWnaive} are obtained by noting that the exact sum rule,
Eq.~\eqref{eq:d00sumrule}, contains the same integral over
cross-sections~\footnote{One can also expand the GMO sum rule Eq.~\eqref{eq:GMO} in
  powers of $M_\pi$.  However, expanding the integrand to match
  Eq.~\eqref{eq:AWnaive} results in a subthreshold expansion
  evaluated at $\nu=M_\pi$ that sits on the radius of convergence of
  the expansion. While truncating this expansion may be a good
  approximation~\cite{Hohler:1983}, it does not result in a rigorous
  chiral expansion.  As current interests lie in the systematic calculation of chiral corrections to the AW sum rule, such an expansion of the GMO sum rule will not be used here.}. Expanding the pion-nucleon coupling constant and
the subthreshold amplitude, $d_{00}^-$, using the results of
Section~\ref{NandC}, leads to
\BE
\ga^2 = 1 \ -\ \frac{2\fpi^2}{\pi} \int_{\mpi}^\infty \frac{d\nu}{\nu^2} k{\ensuremath{\left[ \sigma^{\pi^-p}(\nu) \ -\ \sigma^{\pi^+p}(\nu)\right]}} \ + \ \awd \ ,
\label{eq:AWsumrule}
\EE
with the dimensionless AW discrepancy given by
\begin{align}
\awd &= -1\ +\ 2\fpi^2d^-_{00} \ +\ 4\ga\mpi^2\bar{d}_{18} \ +\ \mathcal{O}(\mpi^4) \label{eq:awd} \\
&= \mpi^2\left(8\left[\bar{d}_1+\bar{d}_2+ 2\bar{d}_5+\frac{\ga\bar{d}_{18}}{2}\right] + \frac{\ga^4}{24\pi^2\fpi^2}\right)\label{eq:awdLECs} \\ & \hspace{1.5cm} -\ M_\pi^3\left(\frac{8+12g_A^2+11g_A^4}{64\pi F_\pi^2 m_N}-\frac{4c_1+g_A^2(c_3-c_4)}{2\pi F_\pi^2}\right) \ +\  \mathcal{O}(M_\pi^4) \ \nonumber.
\end{align}
Values of $d^-_{00}$ and the ${\bar d}_i$ and $c_i$ LECs (with their correlation matrix) may
be obtained from the Roy-Steiner equation analysis of Ref.~\cite{Hoferichter:2015tha}.

In what follows, the $\ompicubed$-corrected sum rule,
Eq.~\eqref{eq:AWsumrule}, will be analyzed using a parametrization of
the total cross-section together with both dependent and independent
determinations of the AW discrepancy.

\section{The AW sum rule confronts experiment}
\label{CALC}

\subsection{Parametrization of total cross-sections}
\label{ptcs}

\begin{figure}
  \centering
  \includegraphics[width = 0.9\textwidth]{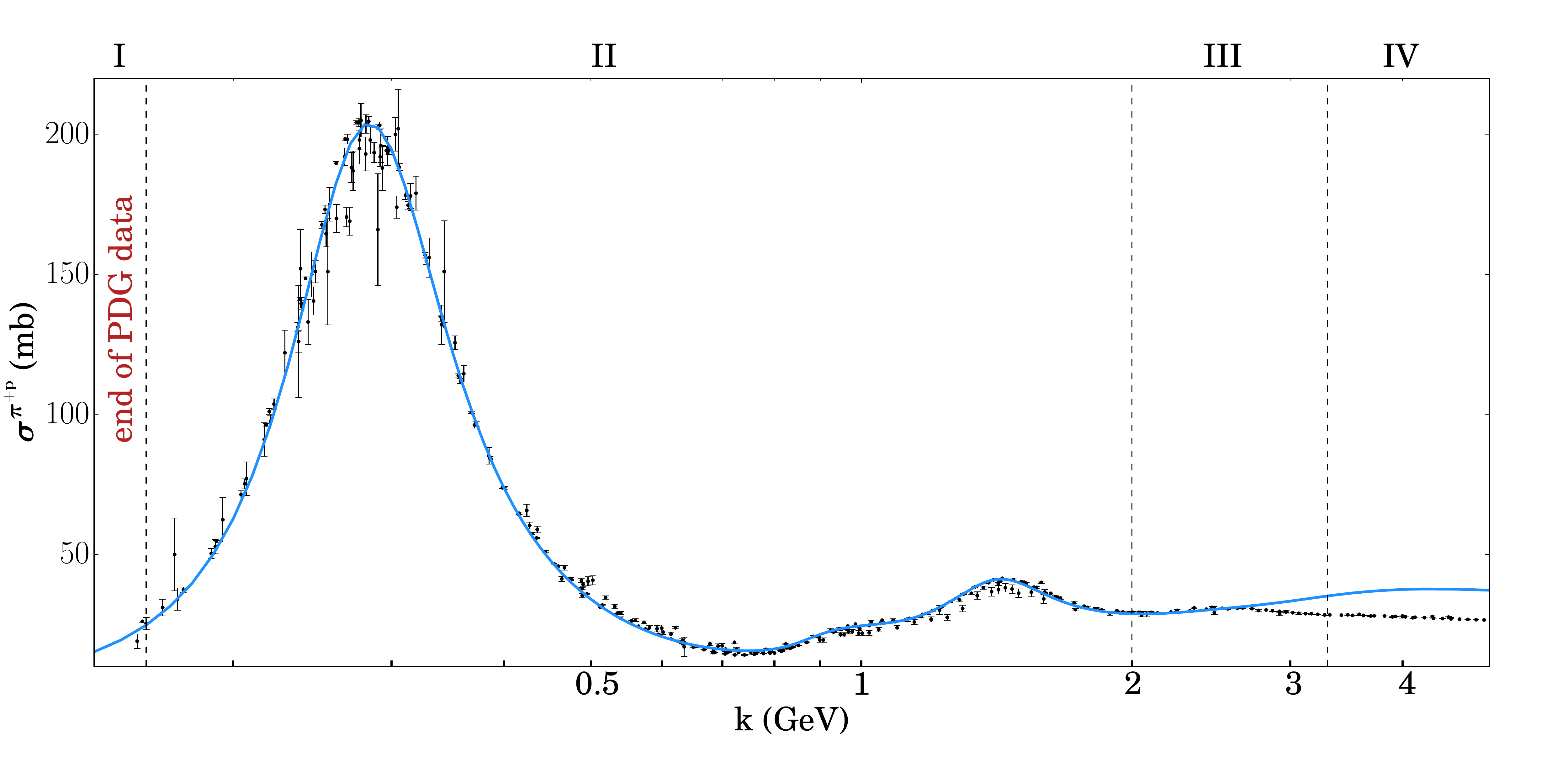}
  \includegraphics[width = 0.9\textwidth]{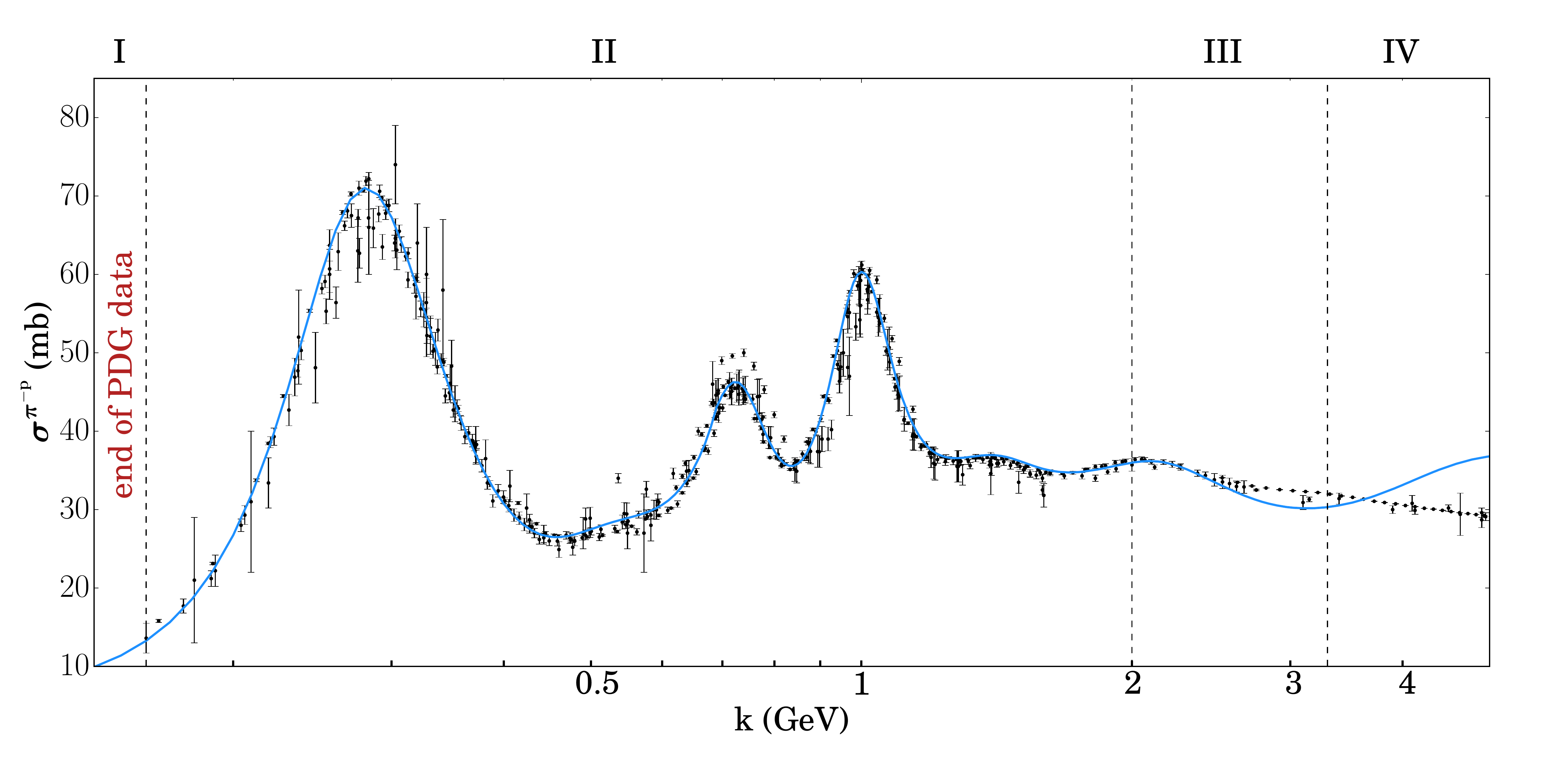}
  \caption{The SAID parameterization~\protect\cite{Workman:2012hx} superposed on the PDG data~\protect\cite{Agashe:2014kda}. This SAID solution is used only
in the resonance region (region II).}
\label{fig:SAIDparameterization}
\end{figure}

In order to confront the chirally-corrected AW sum rule,
Eq.~\eqref{eq:AWsumrule}, with experimental data in a controlled
manner, it is necessary to construct a parametrization of the
cross-section difference $\sigma^-$ of Eq.~\eqref{eq:opticalT} over
all energies.  In what follows, four distinct energy regions are
considered, as outlined in Table~\ref{tab:regions}.  The cross-section
at very-low energies (region I), where there is no PDG
data~\cite{Agashe:2014kda}, is constrained by the effective range
expansion supplemented with the partial-wave expansion, while the
cross-section at very-high energies (region IV) is parametrized
using a Regge-model function fit to PDG $\pi p$ total cross-section
data.  The resonance region (region II) is parametrized by the
recent SAID solution of partial wave fits to $\pi p$
scattering~\cite{Workman:2012hx} (see Fig.~\ref{fig:SAIDparameterization}) while the transition region (region
III) from the resonance region to the Regge region is constructed from
an interpolation of PDG data.

\begin{table}
\centering
  \begin{tabular}{cccc}
  \hline
  \hline
  & & $k$ (GeV) & Source \\
  \hline
  I$_a$ & Threshold & [0.0,0.02] & Effective Range \\
  I$_b$  &    & [0.02,0.16] & PWA~\cite{Workman:2012hx}\\
  II & Resonance & [0.16,2.0] & SAID~\cite{Workman:2012hx}\\
  III & Transition & (2.0,3.3) & PDG~\cite{Agashe:2014kda}\\
  IV & Regge & [3.3,$\infty$] & PDG~\cite{Agashe:2014kda}\\
  \hline
  \hline
  \end{tabular}
  \caption{Regions of the $\pi p$ total cross-sections.
    The distinguishing characteristics  of these regions are the types of data available and the theoretical considerations that enter the parameterization.  PDG data does not exist in region $I$. This region is further divided into $I_a$ where the effective range expansion is valid
    and $I_b$ where parametrizations based on partial wave analyses accurately extend.}
  \label{tab:regions}
\end{table}

\subsubsection*{Region I}

While no experimental data exists for $\sigma^-$ below $k = 0.16$ GeV,
the total cross-section is constrained by various partial wave
analyses and---within its realm of applicability---the effective
range expansion, whose input parameters can be independently determined both
experimentally and from Roy-Steiner-equation
analyses~\cite{Ditsche:2012fv,Hoferichter:2015tha,Hoferichter:2015hva}.

\begin{figure}
  \centering
  \includegraphics[width = 0.9\textwidth]{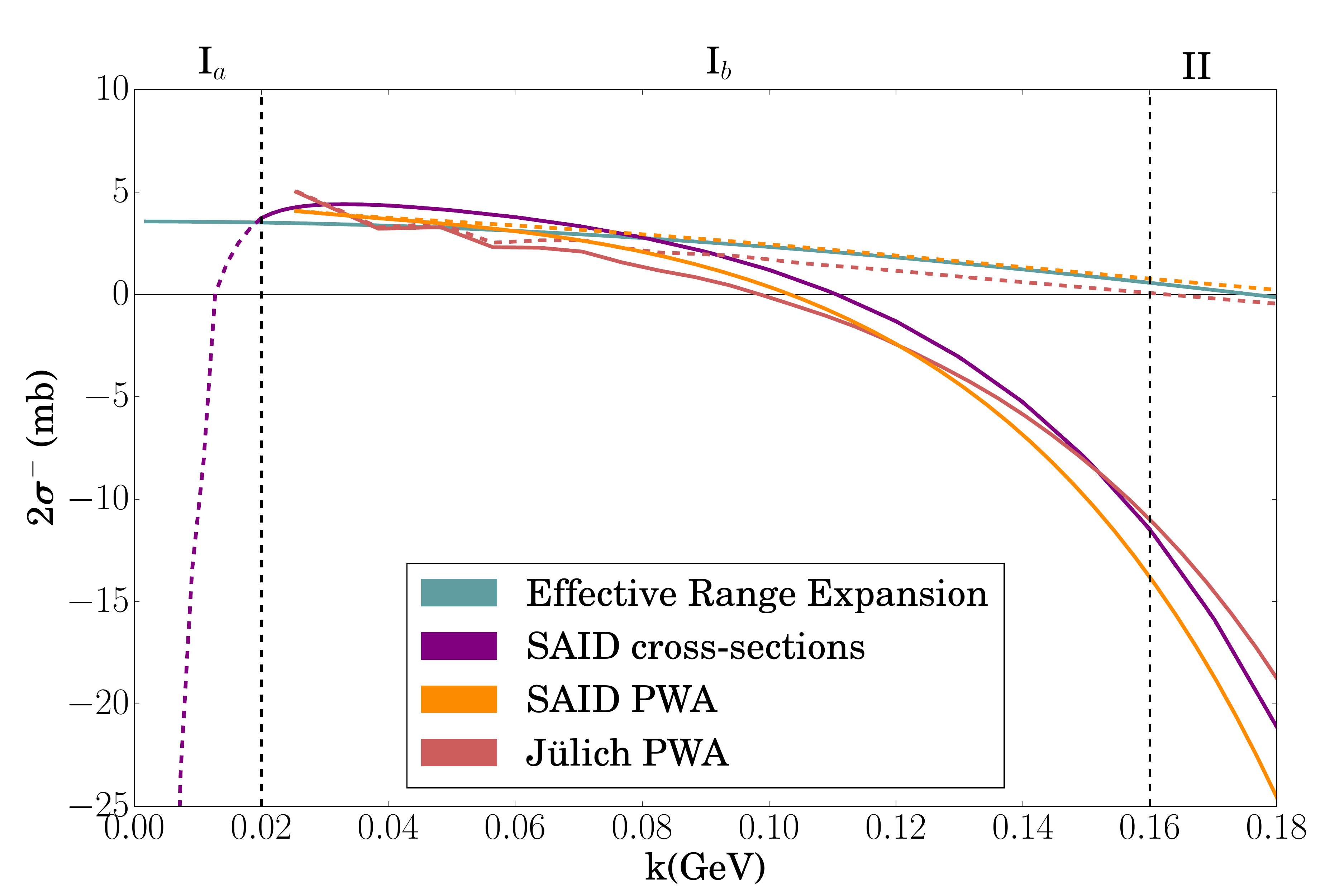}
  \caption{Parameterization of $\sigma^-$ in the
    threshold region with respect to the lab momentum, $k$, of the
    incoming pion.  The dashed and solid lines correspond to S-wave
    and S,P-wave determinations of this quantity.  Clearly, the P-wave
    is an essential contributor.  The low-energy dashed region of the
    SAID WI08 observables solution has not been corrected for Coulomb
    effects and is thus replaced with the effective range expansion in
    this region.}
  \label{fig:Threshold}
\end{figure}

As the lab-frame momentum of the pion approaches zero, the open $\pi^0
n$ channel causes the $\pi^- p$ total cross-section to diverge.
However, the integrated contribution in the region between the $\pi^0
n$ and $\pi^-p$ threshold has been determined to be small~\cite{Ericson:2002md,Baru:2011bw}.  Therefore, isospin invariance is
assumed at $k = 0$.  This allows an effective range expansion of the cross-section,
including the leading momentum dependence, to model the region around
$k = 0$. The first two terms in the effective range expansion are
conventionally parametrized by combinations of isospin even and odd (upper indices $+,-$)
S-wave threshold parameters. In the center-of-mass frame~\cite{Hohler:1983},
\begin{multline}
2\sigma^-(q_{cm}) \ =\  8\pi \Big\lbrack (a_{0+}^-)^2 + 2a_{0+}^+a_{0+}^- +  \\
2q_{cm}^2\left(a_{0+}^-b_{0+}^- + a_{0+}^-b_{0+}^+ + a_{0+}^+b_{0+}^- + \frac{1}{24}(a_1^4 -
a_3^4)\right)\Big\rbrack
\end{multline}
where $q_{cm}$ is the c.m.\ momentum, $a_{0+}^\pm$ ($b_{0+}^\pm$) are
scattering lengths (effective ranges) defined in
Ref.~\cite{Hohler:1983} $a_{1,3}$ are isospin
$\frac{1}{2},\frac{3}{2}$ S-wave scattering lengths, and the
subscripts ${\ell\pm}$ denote total angular momentum states of $j =
\ell \pm \frac{1}{2}$. The relevant isovector and isoscalar scattering
lengths are well known from the spectra of pionic
atoms~\cite{Baru:2011bw}.  In addition, recently an extraction of
scattering lengths and effective ranges for the $\pi p$ system (with
virtual photons removed) has been conducted using Roy-Steiner
equations~\cite{Hoferichter:2015hva}. Using these latter
determinations, one finds (in mb)
\BE
2\sigma^-(q_{cm}) = 3.56(14) - 2q_{cm}^2 86(3) \ ,
\label{eq:EffectiveRange}
\EE
where $q_{cm}$ is expressed in GeV. This parametrization is plotted in
Fig.~\ref{fig:Threshold} together with the results of partial-wave
analyses (PWAs) by the J\"ulich group~\cite{Ronchen:2012eg} and by the
SAID group~\cite{Workman:2012hx}. The region of applicability of the
effective range expansion is less than that suggested by a na{\" i}ve
estimate of its radius of convergence.  Figure~\ref{fig:Threshold}
illustrates that this is due to the influence of the $P_{33}$
($\Delta(1232)$) partial wave, which contributes even at low values of
the pion momentum.  Both the SAID and J\"ulich S-wave determinations
follow the S-wave effective-range expansion throughout this region.
However, the correct structure of $\sigma^-$ is captured only after
the P-wave contributions are included.  Varying the demarcation of
regions I$_a$ and I$_b$ between $k = 0.02$ GeV and $k = 0.08$ GeV is
treated as a means to estimate parameterization-related systematic
uncertainties to the sum rule in the low-energy region.

\subsubsection*{Region IV}

The behaviour of $\sigma^-$ at large momenta (Region IV) is
effectively parametrized by a simple power law decay, consistent with
expectations from the Regge model. This is sufficient for the purposes
of this paper, and $\chi^2$ fitting to PDG data above $k = 3.3$ GeV
gives (in mb)
\BE 2\sigma^-(k) \ =\ 5.76(2) k^{-0.459(1)}  \ ,
\label{eq:reg4fit}
\EE
where again $k$ is in GeV.

\begin{figure}
  \centering
  \includegraphics[width = 0.9\textwidth]{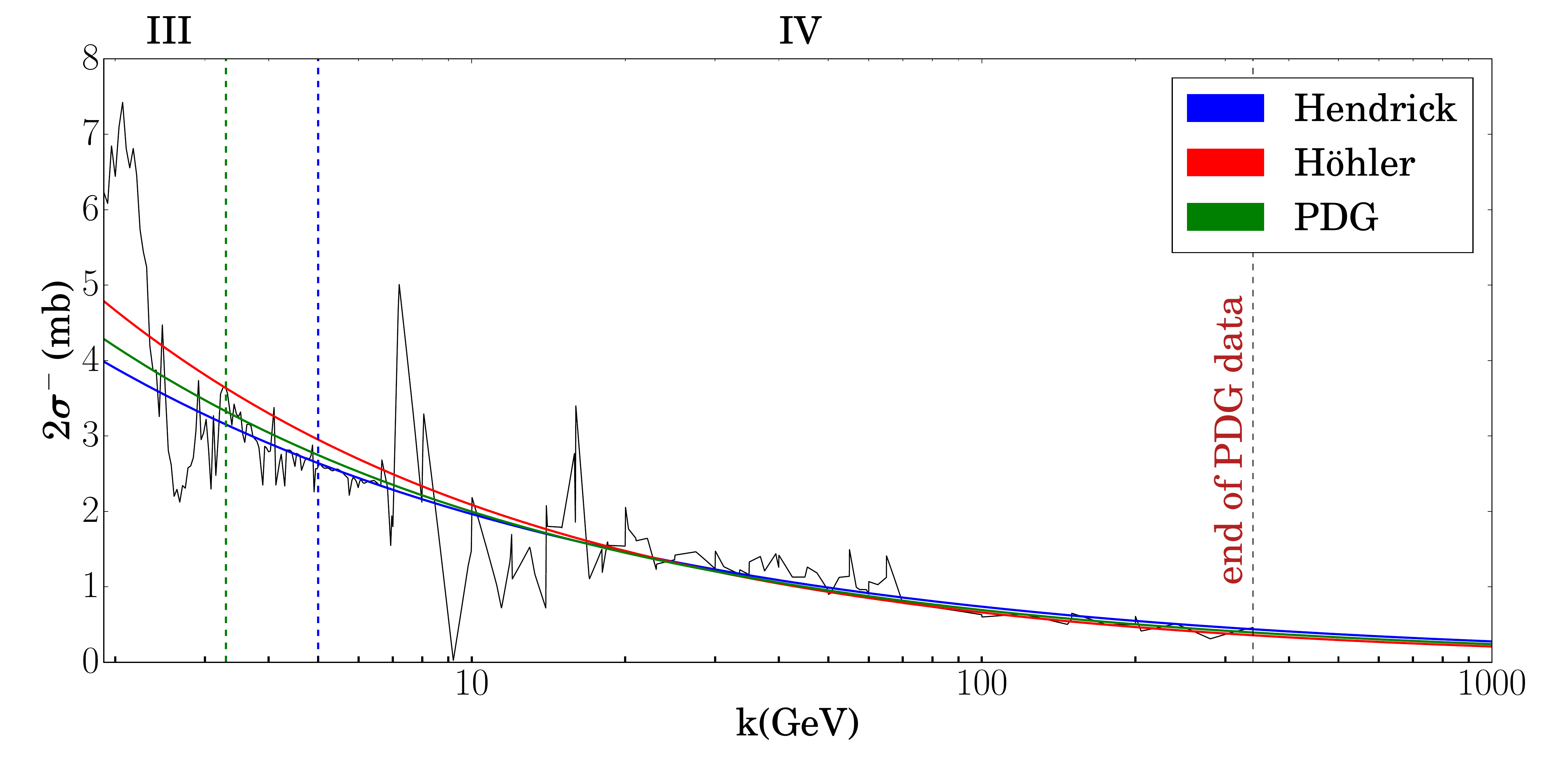}
  \caption{Three distinct (Hendrick\protect\cite{PhysRevD.11.536},
    H\"ohler\protect\cite{Hohler:1983} and
  PDG\protect\cite{Agashe:2014kda}) power-law fits to the high-energy
  region of the sum rule integrand with
respect to lab momentum of the incoming pion, $k$, as in Eq.~\protect\eqref{eq:reg4fit}.
  The dashed lines of corresponding color represent the lower-limit to
  the parametrization's claimed domain of validity.  The sporadic
  black line (ending at the right, black, dashed, vertical line) is a
  raw depiction of the PDG data through $\mathcal{O}(1)$ interpolation
  of the individual $\pi p$ cross-sections contributing to
  $2\sigma^-$ as in the decomposition of Eq.~\eqref{eq:opticalT}.}
  \label{fig:Regge}
\end{figure}

Though other parametrizations of the data have been explored (see
Fig.~\ref{fig:Regge}), the high-energy contributions to the sum rule
are suppressed in the integrand, rendering differences between this simple
parametrization and various other models indistinguishable.  We treat
these alternate fits as a means to estimate systematic uncertainties
to the sum rule in the high-energy region.

\subsection{Testing the parametrization: integral moments}
\label{sec:ExpMoments}

Given the size of the uncertainties due to the integral
parametrization and the GT discrepancy, there are several sources of
uncertainty that are not treated as, comparatively, they constitute
fine structure: isospin violation is not considered, and uncertainties
associated with interpolations of cross-section data are not treated
systematically. One option in the latter case would be to implement a
Gaussian process to interpolate between the $\pi^+$p and $\pi^-$p
cross-section data, propagating the resulting uncertainties to
$\sigma^-$ and to the integral of Eq.~\eqref{eq:AWsumrule}.
Thus, the error bars quoted in this paper are a representation of expectations under reasonable variation of the dominant sources of uncertainty (neither
necessarily gaussian nor defined by a definite probability to encompass the true value).

Calculating the subthreshold amplitudes through evaluation of the
moment sum rules, Eqs.~\eqref{eq:d00sumrule} and
\eqref{eq:momentsumrulesODD}, and comparing results to other
determinations establishes confidence in the parametrization of
$\sigma^-$ developed above.
\begin{table}
\centering
  \begin{tabular}{l|c|c|c|c}
  \hline
  \hline
   & $d_{00}^- \left[\mpi^{-2}\right]$ & $d_{10}^- \left[\mpi^{-4}\right]$ & $d_{20}^- \left[\mpi^{-6}\right]$ & $d_{30}^- \left[\mpi^{-8}\right]$ \\
   \hline
     H\"ohler~\cite{Hohler:1983} & 1.53(2)& -0.167(5)& -0.039(2)& -\\
   \hspace{0.3cm} $\Delta(1232)$ & -0.91+1.17 & -0.18 & -0.04 & - \\
   \hline
   Roy-Steiner Equations~\cite{Hoferichter:2015tha} & 1.41(1) & -0.159(4) & - & - \\
   \hline
   This Paper & 1.50(3) & -0.150(5) & -0.033(2) &-0.0075(8) \\
   \hspace{0.3cm} $\Delta(1232)$ $\delta$-function & 1.9-1.36 & -0.25 & -0.046 & -0.0084 \\
   \hspace{0.3cm} S,P wave & 1.9-0.77 & -0.15 & -0.034 & -0.0089 \\
   \hline
   \hline
  \end{tabular}
  \caption{Calculated values of subthreshold parameters.  Uncertainties represent systematic
    uncertainties associated with alternative parametrizations of regions I and IV and the GT discrepancy, as described in the text.  Listed also are estimates of the $\Delta(1232)$-pole contributions to the moments integrals of Eq.~\eqref{eq:d00sumrule} and Eq.~\eqref{eq:momentsumrulesODD}.
    This paper has constructed two independent estimates of this contribution by (1) saturating the $\pi p$ cross-sections with a $\Delta(1232)$ $\delta$-function and (2) considering the $\pi p$
    cross-sections to be constructed only of S and P partial waves.  The two values stated for $d_{00}^-$ correspond to the $\gpin$ contribution and integral contribution to Eq.~\eqref{eq:d00sumrule}, respectively.}
\label{tab:subthreshold}
\end{table}
Table~\ref{tab:subthreshold} displays the subthreshold parameters as
calculated from (i) the work of H\"ohler~\cite{Hohler:1983} (ii) a
recent analysis of the $\pi p$ amplitude with Roy-Steiner (RS)
equations~\cite{Hoferichter:2015hva}, and (iii) the moment sum rules
using the cross-section parameterization of Section~\ref{ptcs}.  The
uncertainty estimate of $d_{00}^-$ is dominated by the uncertainty in the
value of $\gpin$ stemming from the GT discrepancy.
This explains the order of magnitude larger uncertainties as
compared to the higher moments. To construct this estimate, we have
used the 2\% upper limit expected on the GT
discrepancy as discussed in
Ref.~\cite{Hoferichter:2009gn}. Contributions to the uncertainty
arising from alternative Regge fits or from modifying the threshold
values of the effective-range parameters are comparatively
insignificant, although they are incorporated into the table above.
\begin{figure}
  \centering
    \includegraphics[width = 0.9\textwidth]{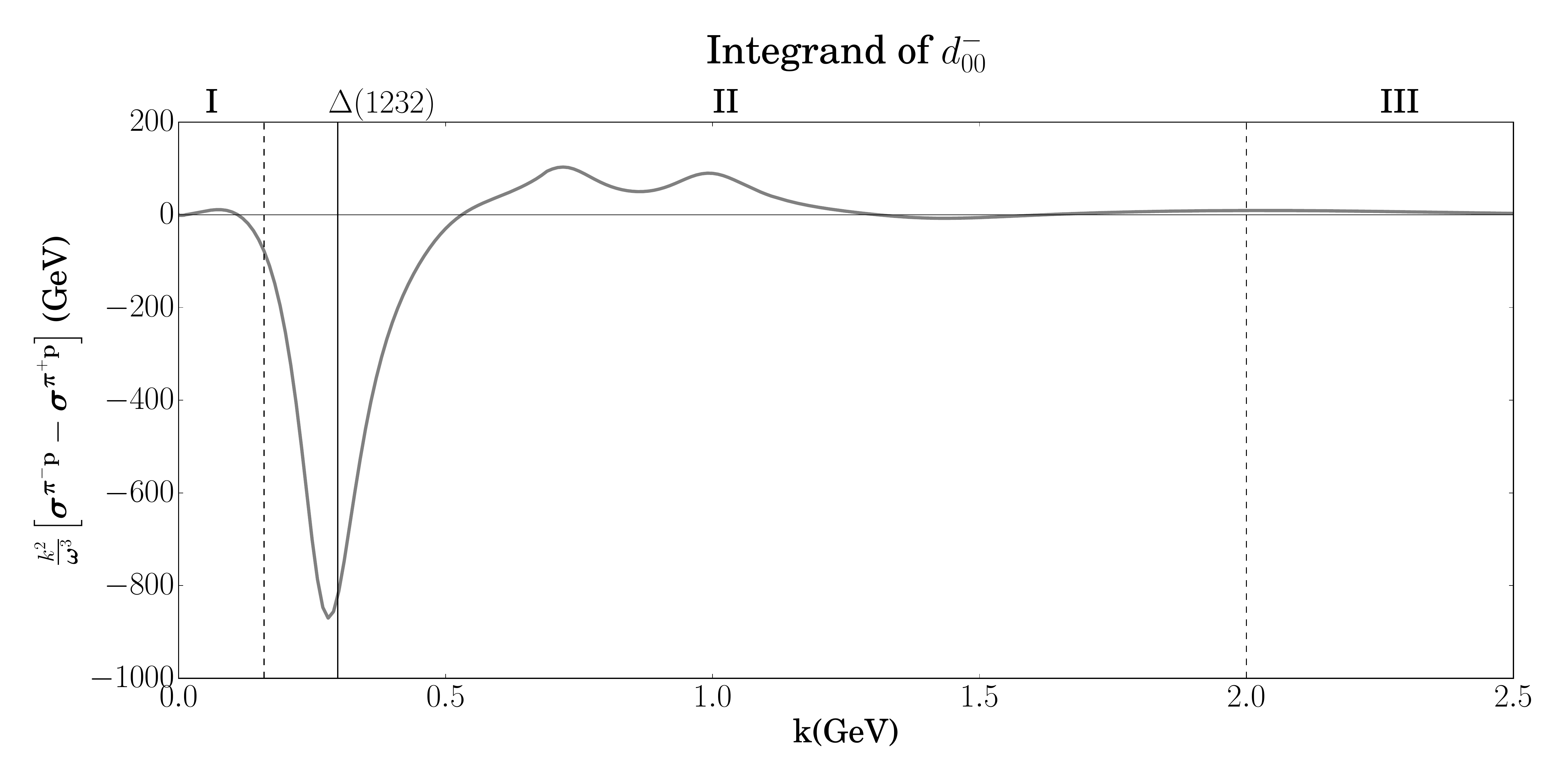}
    \includegraphics[width = 0.9\textwidth]{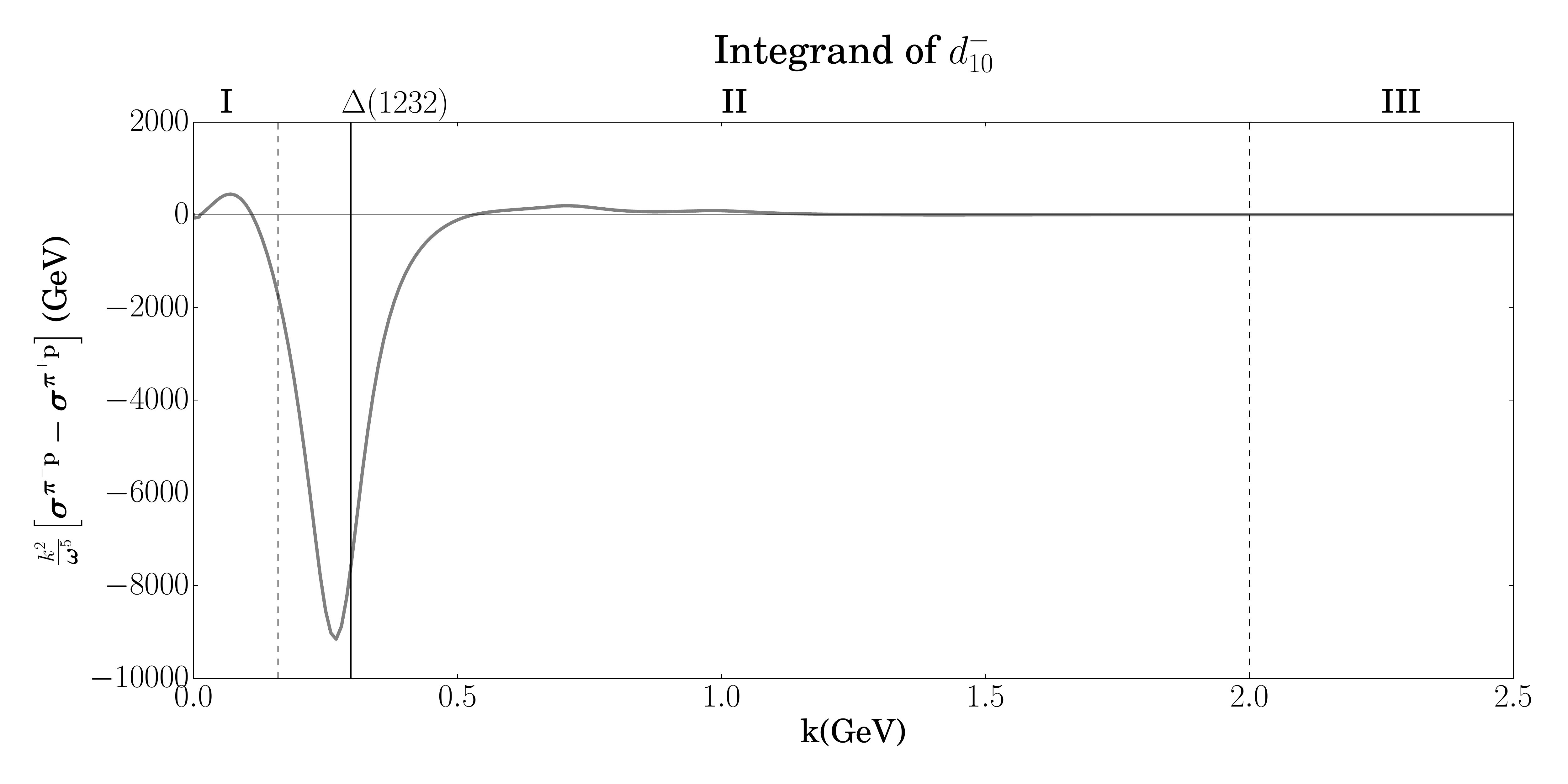}
  \caption{Integrands expressed in the integration variable $k$ associated with the first two subthreshold coefficients.  Note that $k$ is not the variable chosen to express these coefficients in Eqs.~\eqref{eq:d00sumrule} and \eqref{eq:momentsumrulesODD}.  Thus, the solid vertical line is placed at the $\Delta(1232)$ contribution as a pertinent reference.  For higher moments, the integrand tends to increase the influence of the threshold region as well as the influence
   of the $\Delta(1232)$ relative to higher $\Delta$ and $N$ resonances. This is consistent with the numerical findings that the value of subthreshold parameters in Table~\ref{tab:subthreshold} are closely approximated when considering only the $\Delta(1232)$ resonance.}
  \label{fig:d_integrands}
\end{figure}

The higher moments are only sensitive to the cross-section very near
threshold and the $\Delta(1232)$ peak.  Evidently, several of the
coefficients are effectively saturated by the $\Delta(1232)$ resonance
contribution to the sum rule. These observations are illustrated in
Figure~\ref{fig:d_integrands} as well as in
Table~\ref{tab:subthreshold}, where saturation with the $P_{33}$ partial wave
results in a 3\% difference for $d_{20}^-$ and even less for
$d_{10}^-$.  These statements are based on replacing the full PWA of
the resonance region with S and P partial waves only.  Saturation of
the integrand with a $\delta$-function constructed from PDG values for
the $\Delta(1232)$ resonance leads to similar agreement and will be
discussed in greater detail in Section~\ref{EDAWD}~and~\ref{TPP} where, for
comparison with the full continuous parameterization, the integrand is
saturated with N and $\Delta$ resonances of three and four star
PDG significance.  It is reasonable to conclude that beyond these
two coefficients, $d_{00}^-$ and $d_{10}^-$, even the dominant peak of
the $\Delta$ begins to lose its significance in light of the increased
weighting of the threshold region.

We stress that the goal of this section is not to achieve precision
but rather to test the parametrization of the cross-section for
consistency against existing data and theoretical constraints. It is
encouraging that the values of the subthreshold parameters found here
from the moment sum rules are comparable to those found from
independent sources.  The combination of these internal and external
consistencies is taken as license to make use of the
parameterization of Section~\ref{ptcs} in evaluating the
$\mathcal{O}(\mpi^3)$ corrected AW sum rule for $\ga$.

\subsection{Results: the axial-vector coupling constant}
\label{EDAWD}

\begin{table}
  \centering
  \begin{tabular}{l|c|c|c}
  \hline
  \hline
  Eq.~\eqref{eq:awd} & $g_A \, \mathcal{O}(\mpi^2)$ & $\awd $ & \%\\
  \hline
  H\"ohler &  1.282(12) & 0.28(3) & 21.8 \\
  Roy Equations &  1.242(10) & 0.18(3) & 14.5\\
  This Paper  &  1.272(15) \footnotemark & 0.257(36)& 20.2 \\
  \hline
  \hline
  Eq.~\eqref{eq:awdLECs}  & $g_A \, \mathcal{O}(\mpi^2)$ & $\awd $ & \%\\
  \hline
  Roy Equations &  1.255(10) & 0.21(2) & 16.7\\
  \hline
  \hline
  \end{tabular}
  \caption{Calculations of the axial-vector coupling constant and AW discrepancy  (Eq.~\eqref{eq:AWsumrule}) using the subthreshold coefficients of Table~\ref{tab:subthreshold}.
    Uncertainties, as discussed in the text, are estimated from the parameterization, the GT discrepancy \eqref{eq:gtd}, and the truncation of the AW discrepancy beyond
    $\mathcal{O}\left(\mpi^3\right)$. The third column corresponds to the relative contribution of the AW discrepancy to calculations of $g_A$ at this order in $\chi$PT.
  }
  \label{tab:gacalculations}
\end{table}

With a controlled parameterization of the total cross-section over all
energies in hand, the AW sum rule can now be used to determine
$\ga$. Note that $\ga$ appears within the value of the AW discrepancy
itself (see Eq.~(\ref{eq:awd})). Hence, one can treat the AW sum rule
as a non-linear equation for $g_A$, and then use this calculated value
to determine the contribution from $\awd$.  Having done this with the
current parameterization and coefficients from Roy-Steiner equations
leads to the value: $\ga = 1.248 \pm 0.010 \pm \ 0.007 \pm 0.013$, where
uncertainties are from the parametrization of the integral in the sum
rule, the GT discrepancy, and the truncation of the chiral expansion.
Table~\ref{tab:gacalculations} presents the results of this
calculation from Eq.~\eqref{eq:awd} with alternate sets of subthreshold
parameters detailed in Table~\ref{tab:subthreshold} and
Eq.~\eqref{eq:awdLECs} using the LECs of
Ref.~\cite{Hoferichter:2015tha}. The distribution of uncertainties for
these estimates are comparable to that stated above.  In what follows,
we will discuss the sources of uncertainty in some detail.

\footnotetext{This value arises from a \emph{dependent} calculation of $\awd$ in which the integral of Eq.~\eqref{eq:AWsumrule} and the subthreshold parameter $d_{00}^-$ of Eq.~\eqref{eq:awd} are both sourced by the parameterization of Section~\ref{ptcs}.}

The non-linear equation for $g_A$ was solved using
gaussian-approximated, correlated uncertainties for $\bar{d}_1 +
\bar{d_2}$, $\bar{d}_5$, $c_1$, $c_3$, and $c_4$ as well as
uncorrelated uncertainties for $d_{00}^-$, $\bar{d}_{18}$, and the
2012 PDG value of $\fpi$.  These sources of uncertainty are associated
with specific parameterization choices and the GT discrepancy (2\% as
discussed in Ref.~\cite{Hoferichter:2009gn}), and are represented by
the first two numbers of the quoted, partitioned uncertainty for
$g_A$. For the third source of uncertainty, we considered the
truncation of $\awd$ at $\mathcal{O}\left(\mpi^3\right)$. Note that
estimating the truncation uncertainty from, for instance, a number of
order unity times $\left({\mpi}/{4\pi\fpi}\right)^4$, leads to
uncertainties much smaller than those that are quoted.  Instead, the
uncertainty due to truncation is estimated by the consistency of the
analysis in the event that one returns to the dispersion relation,
Eq.~\eqref{eq:MasterDispersion}, and derives a new AW discrepancy. The
alternate expansion that we considered occurs when the pion-mass
dependence of the lab momentum, $k$, appearing in the sum rule
integrand is also expanded in powers of $\mpi$.  While this no longer
arrives at a correction to the chiral-limit AW sum rule, this
resummation allows for an estimate of the influence of neglected
higher order terms.  Using this method leads to an estimated
truncation uncertainty slightly larger than that implied by naive
dimensional analysis.  Included also in this estimate of the
truncation error is the higher-order difference between
Eq.~\eqref{eq:awd} and Eq.~\eqref{eq:awdLECs}.  Together, the three
dominant sources of uncertainty combine to yield the overall
uncertainty stated in Table~\ref{tab:gacalculations}.

\begin{figure}
  \centering
\includegraphics[width=1.01\textwidth]{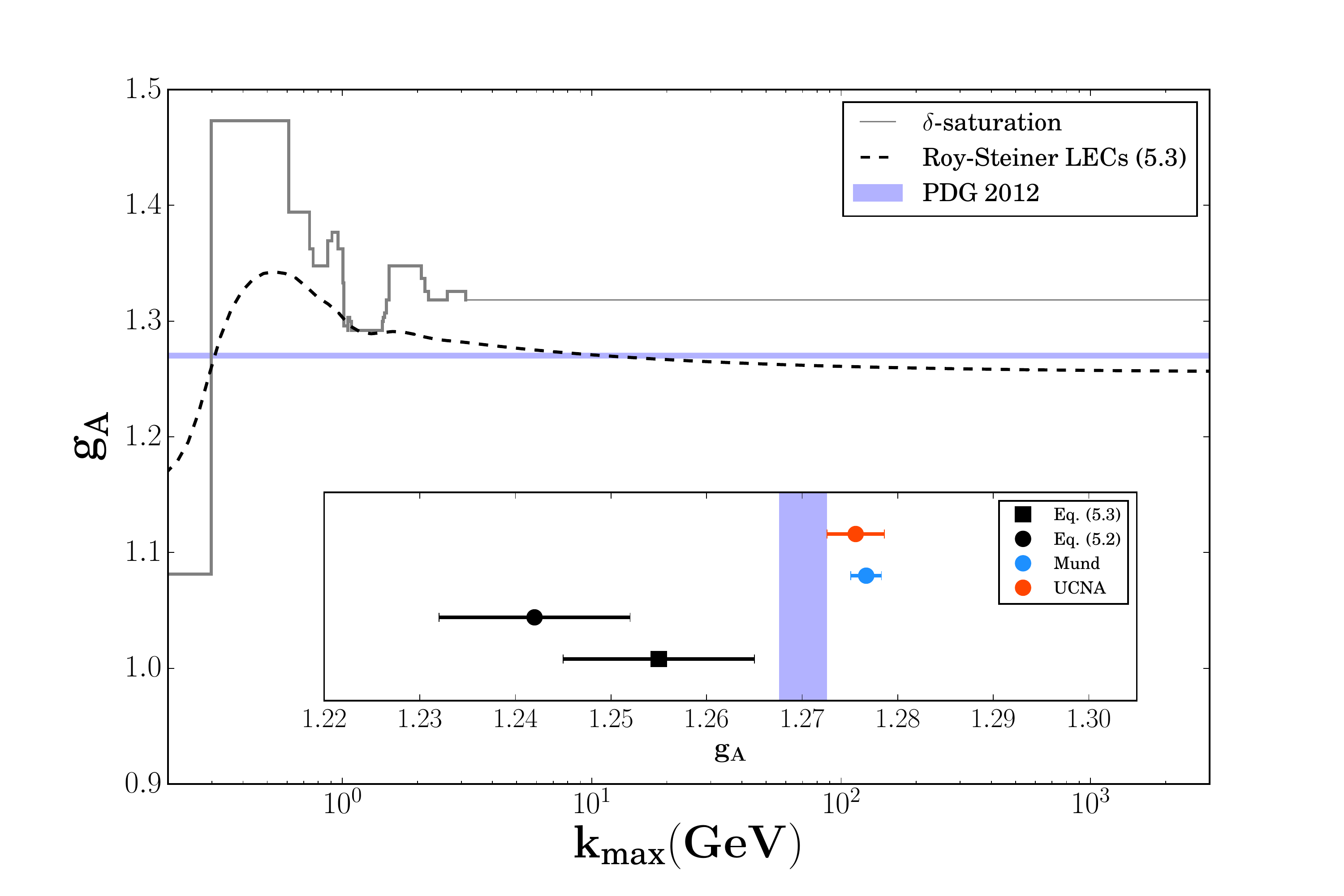}
\caption{Calculated value of $\ga$ with increasing upper bound of the
  integrated total cross-section within the AW sum rule,
  Eq.~\eqref{eq:AWsumrule}.  Results in the main plot are from the
  evaluation of Eq.~\eqref{eq:awdLECs} with the LEC values of
  Ref.~\cite{Hoferichter:2015tha}.  The subplot includes the
  corresponding evaluation with Eq.~\eqref{eq:awd} as well as two
  recent measurements of $\ga$ from neutron $\beta$-decay~\cite{Mund:2012fq,Mendenhall:2012tz} for comparison.  The light gray
  line represents a similar analysis with the total cross-section
  saturated by $\delta$-function resonances~\cite{Beane:2002ud} and
  this paper's evaluation of subthreshold parameters. The light-blue
  band is the 2012 PDG value for $\ga$.}
  \label{fig:ga_kmax}
\end{figure}

Focusing on the independent evaluations which rely on recent
Roy-Steiner calculations of $d_{00}^-$ and the correlated LECs, one
finds that the two $\mathcal{O}(\mpi^3)$ evaluations of $\ga$ are
internally consistent.  More importantly, one finds that the magnitude
and sign of $\awd$ are in agreement with the experimental observation
that the value of $\ga$ is approximately 25\% larger than its chiral
limit value.  In the next section, we will examine this symmetry
breaking in greater detail.

\subsection{The physical picture}
\label{TPP}
The AW sum rule is a constraint on the flow of null-plane, axial-vector
charge between the nucleon and all other states that the nucleon can
transition to through the emission or absorption of a pion. The
transitions can occur only because the charges are not conserved (they
depend on $x^+$) and therefore they are able to mediate the energy
transfer that is necessary for the processes to take place.  Of
course, physically, the non-conservation of the null-plane
axial-vector charge signals spontaneous chiral symmetry breaking. This
picture is, strictly speaking, correct only in the chiral limit and
therefore in this case the deviations of $\ga$ from unity are a
measure of spontaneous symmetry breaking.  As we have seen here,
$\chi$PT allows the quantitative inclusion of corrections to this
picture due to non-vanishing light-quark masses via $\Delta_{AW}$.  An
intuitive visual representation of the sum rule gives the value of
$\ga$ as a function of the upper value of the integration momentum ($k_{\text{max}}$) as
it is increased from zero to its asymptotic value.  (See
Figure~\ref{fig:ga_kmax}.) As one sees in the plot, methodically
adding states of higher energy under the integral (increasing $k_{\text{max}}$) adds and subtracts
chiral charge, depending on the intermediate state.

When the chiral-limit sum rule (Eq.~\eqref{eq:AWcl}) is expressed in terms of physical quantities to produce the leading, $\mathcal{O}(\mpi^0)$  contribution (Eq.~\eqref{eq:AWnaive}), the axial-vector
coupling constant at $k_{max} = M_\pi$ is exactly 1. With the introduction of chiral corrections, this value is shifted to $1+\Delta_{AW}$.  Once the
chiral symmetry is spontaneously broken, intermediate states that
transition to the nucleon via the non-conserved axial-vector charge
can and do appear.  In the interest of gaining understanding of the
weightings associated with these states, depicted by the evolution of
the integral in Figure~\ref{fig:ga_kmax}, the integrand can be modeled
with a finite number of known resonances which couple strongly to the
pion-nucleon system.  This process, $\delta$-saturation, was carried
out in Ref.~\cite{Beane:2002ud}, where the cross-sections
participating in the AW sum rule were approximated by
$\delta$-functions of the appropriate N and $\Delta$ resonances using
the chiral-limit form of the sum rule. Here, this exercise is
repeated, but including the effect of the AW discrepancy. Figure~\ref{fig:ga_kmax} shows that the delta functions lead to a series of
step functions in the calculation of $g_A$ that, as expected,
qualitatively track the curvature of the actual integrand obtained
from the parameterization of cross-sections.

While the $\delta$-saturation of Ref.~\cite{Beane:2002ud} neglected
the AW discrepancy, the analysis resulted in an evaluation of $g_A
\simeq 1.26$---a value surprisingly close to experiment, albeit with
no measure of uncertainty. Saturating the AW sum rule using the same
set of resonances but with Breit-Wigner line shapes and a threshold
region as discussed in Section~\ref{ptcs} yields the value $\ga \simeq
1.27$ (with $\Delta_{AW}=0$).  With the now-improved understanding of
the chiral corrections to the AW sum rule, these past successes of
$\delta$-saturated models may seem more fortuitous than
illuminating. However, both this ``leading-order'' agreement and the
qualitative agreement of Figure~\ref{fig:ga_kmax} indicates that models of
pion-nucleon scattering, and more generally of the nucleon null-plane
wave-function, that implement a finite number of resonances, provide
an approximate description that could prove useful for modeling the
internal axial structure of the nucleon.

The subplot of Figure~\ref{fig:ga_kmax}, provides a comparison between
sum rule determinations of $\ga$ and current experimental measurements
of the coupling constant.  According to the 2012 PDG review, $\ga =
1.2701(25)$.  Recent experimental measurements of the neutron
$\beta$-decay asymmetry parameter gives $\ga =
1.276(3)$~\cite{Mund:2012fq}\cite{Mendenhall:2012tz}.  While the
uncertainties that arise in the AW sum rule determination of $\ga$
presented in this paper are not particularly aggressive (claiming high precision), the results
bring the sum-rule determination of $\ga$ into consistency with
current measured values of $\ga$ and emphasize the physical mechanism
of QCD that is responsible for the axial-vector charge's deviation
from unity.

\section{Conclusions}
\label{conc}

The AW sum rule is a unique signature of chiral symmetry and its
breaking in QCD as its validity resides in both the algebraic content
of chiral symmetry, which guarantees the convergence of the sum rule,
and the dynamical content of chiral symmetry, which allows the
systematic inclusion of light-quark mass effects. In this paper, it
has been shown how, using results of $\chi$PT, the chiral limit sum
rule may be systematically extended to include corrections up to
$\mathcal{O}(M_\pi^3)$. In addition, the introduction of the AW
discrepancy allows a non-unique but useful means of separating the
contributions to the deviation of $\ga$ from unity into distinct parts
that arise from spontaneous and explicit chiral symmetry breaking.

While the calculation presented here is, by construction, independent
of experimental measurements of $\ga$, the parameterization we have
established may be useful beyond the determinations of $\awd$ and
$\ga$ provided here. Considering the current precision of $\ga$
measurements, it is reasonable to consider rearranging the
$\ompicubed$ sum rule to take the value of $\ga$ as experimental input
for a determination of LECs.  For example, recall that $\awd$ may be
expressed in terms of a linear combination of LECs of the $\pi p$
system (Eq.~\eqref{eq:awdLECs}).  Thus, $\ga$ is a physical quantity
with direct dependence on the LEC correlation matrix---a now essential
piece of any LEC extraction.  Similarly, the AW sum rule may be used
to constrain the LEC $\bar{d}_{18}$, which parametrizes the GT
discrepancy, a significant source of uncertainty in many calculations,
including those of $\gpin$.  Whether the AW sum rule (with the
correction of $\awd$) will provide significant constraints on such
LECs will be left as a question for future research.

\acknowledgments{ We thank M.~Hoferichter for valuable conversations
  and R.~Workman for providing useful information regarding the SAID group
  analyses.  The work of SRB was supported in part by the U.S.
  National Science Foundation through continuing grant PHY1206498 and
  by the U.S. Department of Energy through Grant Number DE-SC001347.
  The work of NMK was supported in part by the University of
  Washington's Nuclear Theory Group and by the Seattle Chapter of the
  Achievement Rewards for College Scientists foundation.  }

\bibliographystyle{KP}
\bibliography{bibi}

\end{document}